\documentclass[%
reprint,
superscriptaddress,
 amsmath,amssymb,
 pre,
]{revtex4-1}

\usepackage{graphicx}
\graphicspath{{./figs/}}
\usepackage{dcolumn}
\usepackage{bm}

\begin{document}

\preprint{APS/123-QED}

\title{Transient Growth of Ekman-Couette Flow}

\author{Liang Shi}
\email{liang.shi@ds.mpg.de}
\affiliation{%
  Max Planck Institute for Dynamics and Self-Organization (MPIDS), 37077 G\"ottingen, Germany
}
\affiliation{%
  Institute of Geophysics, University of G\"ottingen, 37077 G\"ottingen, Germany
}

\author{Bj\"orn Hof}
\email{bhof@ist.ac.at}
\affiliation{%
  Max Planck Institute for Dynamics and Self-Organization (MPIDS), 37077 G\"ottingen, Germany
}
\affiliation{%
  IST Austria, 3400 Klosterneuburg, Austria
}

\author{Andreas Tilgner}
\email{andreas.tilgner@physik.uni-goettingen.de}
\affiliation{%
  Institute of Geophysics, University of G\"ottingen, 37077 G\"ottingen, Germany
}

\date{\today}

\begin{abstract}
Coriolis force effects on shear flows are important in geophysical and 
astrophysical contexts. 
We here report a study on the linear stability and the transient energy growth of the 
plane Couette flow with system rotation perpendicular to the shear direction. 
External rotation causes linear instability. At small rotation rates, the onset 
of linear instability scales inversely with the rotation rate and the optimal transient 
growth in the linearly stable region is slightly enhanced, $\sim \text{\text{Re}}^2$. 
The corresponding optimal initial perturbations are characterized by roll structures inclined 
in the streamwise direction and are twisted under external rotation. At 
large rotation rates, the transient growth is significantly inhibited and hence linear stability 
analysis is a reliable indicator for instability.

\end{abstract}

\maketitle

\section{Introduction}
\begin{figure}[!h]
  \centering
  \includegraphics[width=0.4\textwidth]{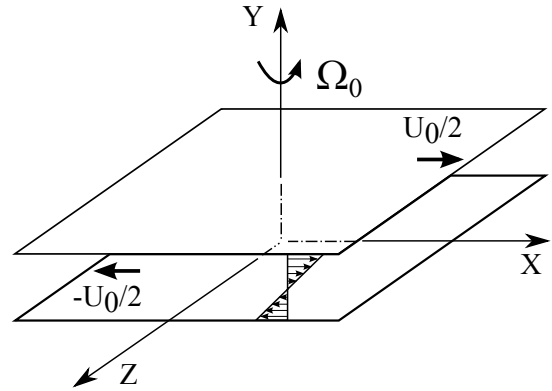}
  \caption{Schematic of Ekman-Couette flow. The top and bottom walls slide  
    both with velocity $U_0/2$ but in opposite directions along X axis. The whole 
    setup rotates at a speed of $\Omega_0$ around the Y axis. The velocity 
    profile corresponds to the base flow of plane Couette flow, $(U_0/2\cdot y, 0,0)$.}
  \label{fig:eckmannCouette}
\end{figure}

Ekman-Couette flow represents the flow between two sliding parallel walls, where 
the whole setup is subject to external rotation around the axis perpendicular to the walls.
Fig.~\ref{fig:eckmannCouette} shows schematically the geometry of the flow. 
In the extreme cases, the flow becomes either plane Couette flow (pCf, 
if without rotation) or two well separated Ekman layers for large rotation rates. Because of 
the theoretical importance and the practical generality in planetary systems, these two 
canonical shear flows have both received enormous attention for the last decades and pCf 
under spanwise system rotation has also been widely studied
~\cite{Rincon_aa2007, Tsukahara_jfm2010}. However, 
little work has been done to study the Ekman-Couette flow. Hoffmann \textit{et. al.} 
~\cite{HoffmannChen_jfm1998} studied the secondary and tertiary flow states 
in Ekman-Couette flow while Ponty \textit{et. al.}~\cite{PontySoward_jfm2003} investigated the 
onset of thermal convection between two shearing plates under the influence 
of external oblique rotation. Both studies focused mainly on the regime of 
moderate to large external rotation. 
The instability at small rotation has been yet payed little attention to. 
Since experimental flows on Earth are mostly subject to weak external rotation and the Earth's rotation has 
been reported to have measurable influences in many other 
flows~\cite{Draad_jfm1998,Brown_pof2006,Boisson_epl2012,TrianaLathrop_jgr2012}, 
the influence of weak system rotation on the Couette flows will be here specially studied.

In this paper we present a study on the linear stability 
and the transient energy growth exploring a wide parameter space in Ekman-Couette 
flow. This work is theoretically interesting and is also motivated by recent 
conflicting results~\cite{Balbus_nature2011,
JiGoodman_nature2006,SchartmanGoodman_aa2012,PaolettiLathrop_prl2011,Avila_prl2012} 
in astrophysical rotating flows, on whether turbulence in cold accretion disks can 
arise via hydrodynamic instabilities. The Ekman layers introduced by the top and 
bottom end walls in experimental Taylor-Couette setups influence remarkably 
the bulk flow and make the flow rather complicated. Besides, the Earth's rotation gives rise to 
another component of rotation, perpendicular to the rotation axis of the cylinders. At 
high \text{\text{Re}}, its effects may become non-negligible, except that the rotation axis of the cylinders 
aligns with the one of Earth's rotation. We here choose the simplest geometry 
to study the influence of the Ekman layer on the linearly stable flows. We find
that in pCf an infinitesimal external rotation causes linear instabilities. 

This paper is structured in the following way. The linearized Ekman-Couette problem is 
formulated mathematically in Section 2, followed by the linear stability analysis in 
Section 3. We finally study the transient energy growth in Section 4. 

\section{Problem formulation}
\begin{figure}[!h]
  \centering
  \vspace{16pt}
  \includegraphics[width=0.4\textwidth]{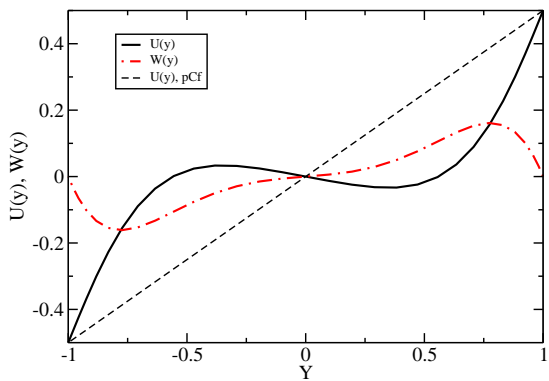}
  \caption{(Color online) Base velocity profiles at $\text{Re}=1000, \Omega = 50$ (solid and dash-dotted lines) and at 
    $\text{Re}=1000, \Omega = 0$ (dashed line). }
  \label{fig:basVel}
\end{figure}

Considering that the fluid is incompressible, the governing equations of the fluid motion 
are the Navier-Stokes equations,
\begin{equation}
  \partial_{t}\textbf{u}+\textbf{u}\cdot\nabla\textbf{u} + 2\Omega_0 
  \textbf{e}_y\times\textbf{u} =-\frac{1}{\rho}\nabla
  p+\nu\Delta\textbf{u},\quad \nabla\cdot\textbf{u}=0.
  \label{eq:NS}
\end{equation}
where $\textbf{u}(\textbf{x},t)$ is the flow velocity field and $p(\textbf{x},t)$ 
is the pressure field. By taking the half gap distance between two plates $D/2$ as 
the length unit and $D/(2U_0)$ as the time unit, 
\begin{equation*}
  l=l'\cdot D/2, \quad t=t'\cdot D/(2U_0), \quad \textbf{u} 
  = \textbf{u}'\cdot U_0, \quad p = p'\cdot \rho U_0^2,
\end{equation*}
we obtain the nondimensional form of Eq.~\ref{eq:NS}, 
\begin{equation}
  \partial_{t}\textbf{u}+\textbf{u}\cdot\nabla\textbf{u} + \frac{1}{\text{Ro}}\cdot 
  \textbf{e}_y\times\textbf{u} =-\nabla
  p+\frac{1}{\text{Re}}\cdot\Delta\textbf{u},\quad \nabla\cdot\textbf{u}=0.
  \label{eq:nondimNS}
\end{equation}
with the \text{Re}ynolds number and the Rossby number 
\begin{equation*}
  \text{Re}=\frac{U_0D}{2\nu}, \quad \text{Ro}=\frac{U_0}{\Omega_0 D}
\end{equation*}
Note that the nondimensional symbols in Eq.~\ref{eq:nondimNS} are omitted.
We define another nondimensional parameter, the rotation number 
$\Omega=\frac{\Omega_0 D^2}{\nu}$. Here, $\text{Ro} = \frac{2\text{Re}}{\Omega}$.
Considering the boundary conditions and the symmetry property about the plane 
$y=0$, the base velocity profile has the form of $[U(y),0,W(y)]$. We 
introduce the complex function $Z(y)=U(y)+iW(y)$ and yield
\begin{equation}
  Z(y)=\frac{1}{2}\frac{e^{i\gamma y}-e^{-i\gamma y}}{e^{i\gamma}-e^{-i\gamma}},
\end{equation}
with $\gamma=\sqrt{\frac{\text{Re}}{\text{Ro}}}\frac{1+i}{\sqrt{2}}$. 
Fig.~\ref{fig:basVel} displays the base velocity profile at $\text{Re}=1000$, with and 
without rotation, respectively. At $\Omega = 50$, the external rotation distorts 
qualitatively the base flow such that the inflection points appear in the profiles. 

To study the linear stability and transient dynamics of the base flow, 
we decompose the velocity field as $\textbf{u}=\textbf{u}_{pert}+\textbf{U}_{base}$, 
where $\textbf{U}_{base}=[U(y),0,W(y)]$. Let $v$ and $\eta$ 
denote the perturbation of the wall-normal velocity and vorticity. 
By taking the curl once and twice, respectively, of Eq.~\ref{eq:nondimNS} and 
then projecting into the Y direction, we obtain the linearized 
equations for the perturbation variables $(v,\eta)$,

\begin{equation}
  \begin{aligned}
    \partial_t\nabla^2 v + (U\partial_x+W\partial_z)\nabla^2 v - W''\partial_z v - U''\partial_x v =& \\
    \frac{1}{\text{Re}}\nabla^4 v - \frac{1}{\text{Ro}}\partial_y\eta, & \\
    \partial_t\eta + (U\partial_x+W\partial_z)\eta + U'\partial_z v - W'\partial_x v =&\\
    \frac{1}{\text{Re}}\nabla^2 \eta + \frac{1}{\text{Ro}}\partial_y v. &
  \end{aligned}
  \label{eq:pertNS}
\end{equation}

In this paper we focus on the following modal perturbation, 
\begin{equation*}
  v=\hat{v}(y,t)e^{i(\alpha x+\beta z)}, \quad \eta=\hat{\eta}(y,t)e^{i(\alpha x+\beta z)},
\end{equation*}
where $\alpha$ and $\beta$ are the wavenumbers in $X$- and $Z$- direction, respectively. 
By inserting into Eq.~\ref{eq:pertNS}, we have the modal equations,

\begin{equation}
  \begin{aligned}
    \partial_t\hat{\nabla}^2\hat{v} =& -i(U\alpha+W\beta)\hat{\nabla}^2\hat{v} + i(U''\alpha+W''\beta)\hat{v} \\
    & + \frac{1}{\text{Re}}\hat{\nabla}^4 \hat{v} - \frac{1}{\text{Ro}}\partial_y\hat{\eta}, \\
    \partial_t\hat{\eta} =& - i(U\alpha+W\beta)\hat{\eta} - i(U'\beta - W'\alpha)\hat{v} \\
    & + \frac{1}{\text{Re}}\hat{\nabla}^2 \hat{\eta} + \frac{1}{\text{Ro}}\partial_y \hat{v},
  \end{aligned}
  \label{eq:fourierNS}
\end{equation}
with $\hat{\nabla}^2=\partial_y^2-(\alpha^2+\beta^2)$.

Through the Chebyshev spectral discretization in the spatial 
direction~\cite{ReddyHenningson_jfm1993}, the above partial differential equations is 
transformed into a linear system $\partial_t \hat{\textbf{v}} = -iL\hat{\textbf{v}}$, 
where $\hat{\textbf{v}}=
[\hat{v}, \hat{\eta}]$. The linear stability and transient 
growth is then calculated by the eigenvalues and eigenvectors of the linear operator $L$, 
which is computed in this paper by the subroutines in the {\tt LAPACK} library. 
The accuracy and convergency of the method have been verified against the results in
~\cite{ReddyHenningson_jfm1993}.

\section{Linear instability}
\begin{figure}[!h]
  \centering
  \vspace{10pt}
  \includegraphics[width=0.4\textwidth]{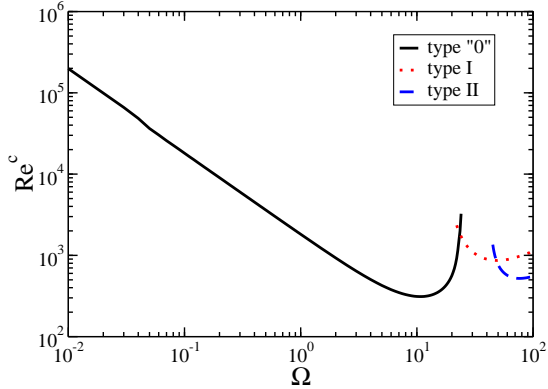}\\
  \caption{(Color online) Critical \text{Re}ynolds number $\text{Re}^c$ as a function of $\Omega$. 
    For $\Omega < 5$, the critical \text{Re} scales inversely with the system rotation, $\text{Re}^c\simeq 1800\cdot\Omega^{-1}$.}
  \label{fig:Rec}
\end{figure}

\begin{figure}[!h]
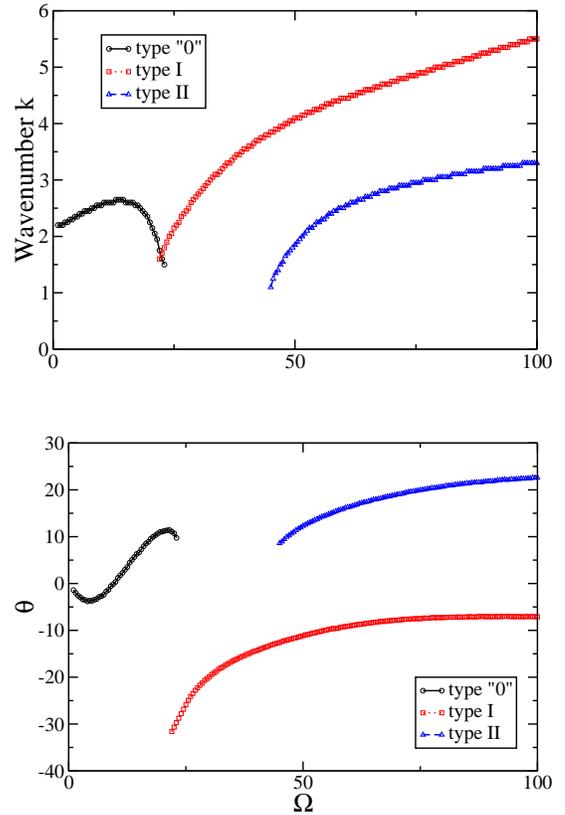

  \centering
  \includegraphics[width=0.4\textwidth]{k_c_v2.eps}\\[15pt]
  \includegraphics[width=0.4\textwidth]{theta_c_v2.eps}\\
  \caption{(Color online) The variation with $\Omega$ of the wavenumber $k=\sqrt{\alpha^2+\beta^2}$ (top) 
    and the angle $\theta$ (bottom).}
  \label{fig:wavnumTheta}
\end{figure}

The inflection points in the base profile hint that the Ekman-Couette flow 
may be linearly unstable. Thus we first inverstigate the linear instability of the flow. 
The range of parameters under study is $\text{Re} \in [100,300000]$ 
and $\Omega \in [0,100]$. A bisection method is employed to find the critical 
curve $\text{Re}^c(\Omega)$, seperating the linearly stable and unstable regions. 
The results are shown in Fig.~\ref{fig:Rec}. At small $\Omega$ ($\Omega<5$), 
the linear instability is here referred to as type ``0'' and the critical \text{Re}ynolds 
number $\text{Re}^c$ is found to scale with $\Omega$ as $\text{Re}^c (\Omega) \simeq 1800\cdot\Omega^{-1}$.
Therefore, as $\Omega \rightarrow 0$, $\text{Re}^c \rightarrow \infty$,
which is consistent with the linear stability of plane Couette flow ($\Omega = 0$) 
at any $\text{Re}$
~\cite{OrszagKells_jfm1980,DrazinReid_1981}. As $\Omega$ is increased, 
we recover the type I and type II instabilities previously found in Ekman layer flow~
\cite{Lilly_1966}. The corresponding wavenumbers are shown in Fig.~\ref{fig:wavnumTheta}. 
Here the wavenumber $k=\sqrt{\alpha^2+\beta^2}$ and the angle $\theta = -\text{arctan}(\alpha/\beta)$, 
where $\theta$ is the angle between the wavevector $\textbf{k}$ and the $Z$ axis. 
The negative sign indicates anti-clockwise direction.
As shown in the Fig.~\ref{fig:wavnumTheta}, type I instability is characterized by a large 
wavenumber and a negative angle while type II has a smaller wavenumber and a positive angle. 
The results agree very well with the ones previously reported 
in~\cite{HoffmannChen_jfm1998, PontySoward_jfm2003}.

\section{Transient growth}
\begin{figure*}[!ht]
  \centering
  \includegraphics[width=0.4\textwidth]{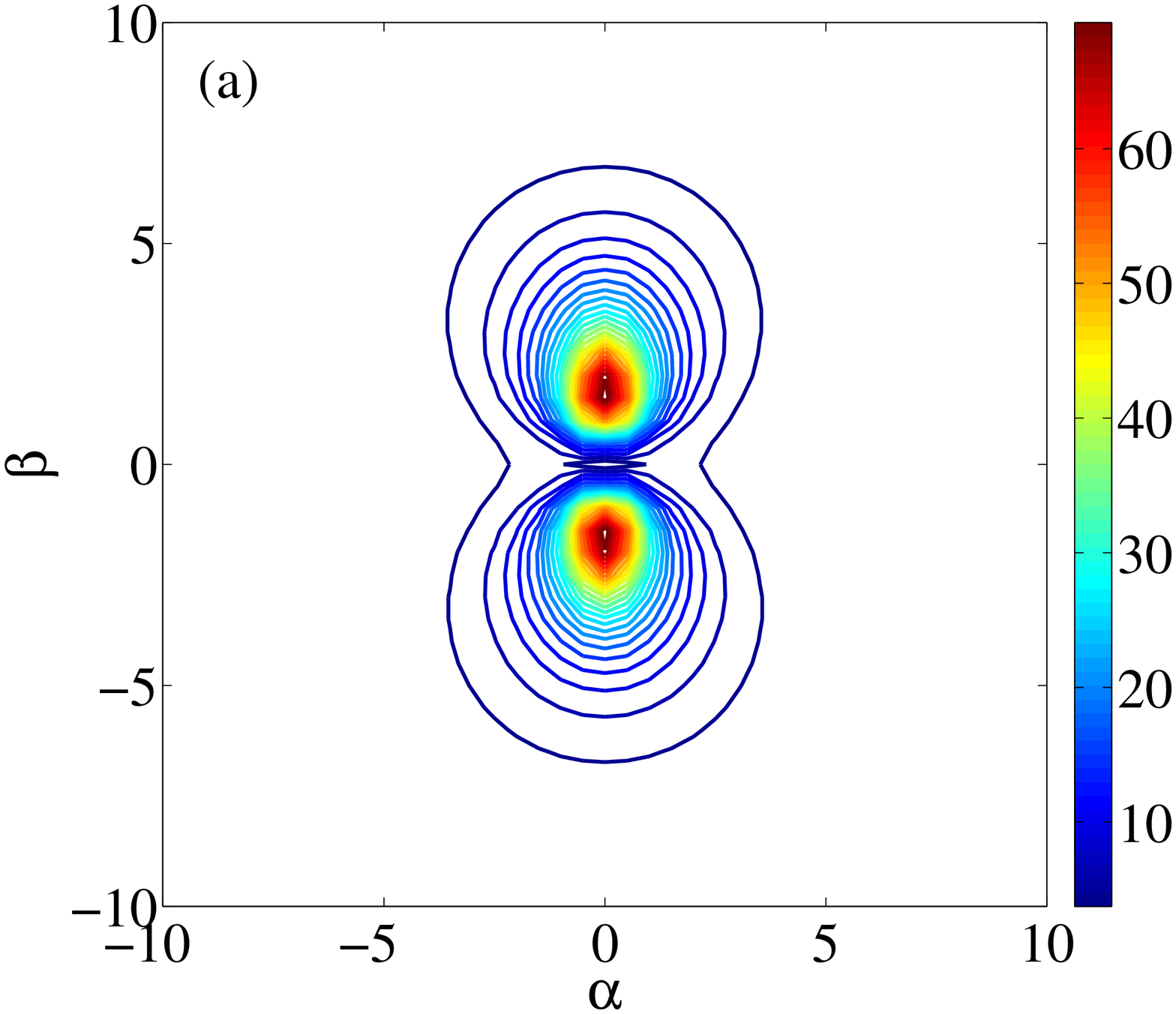}
  \includegraphics[width=0.4\textwidth]{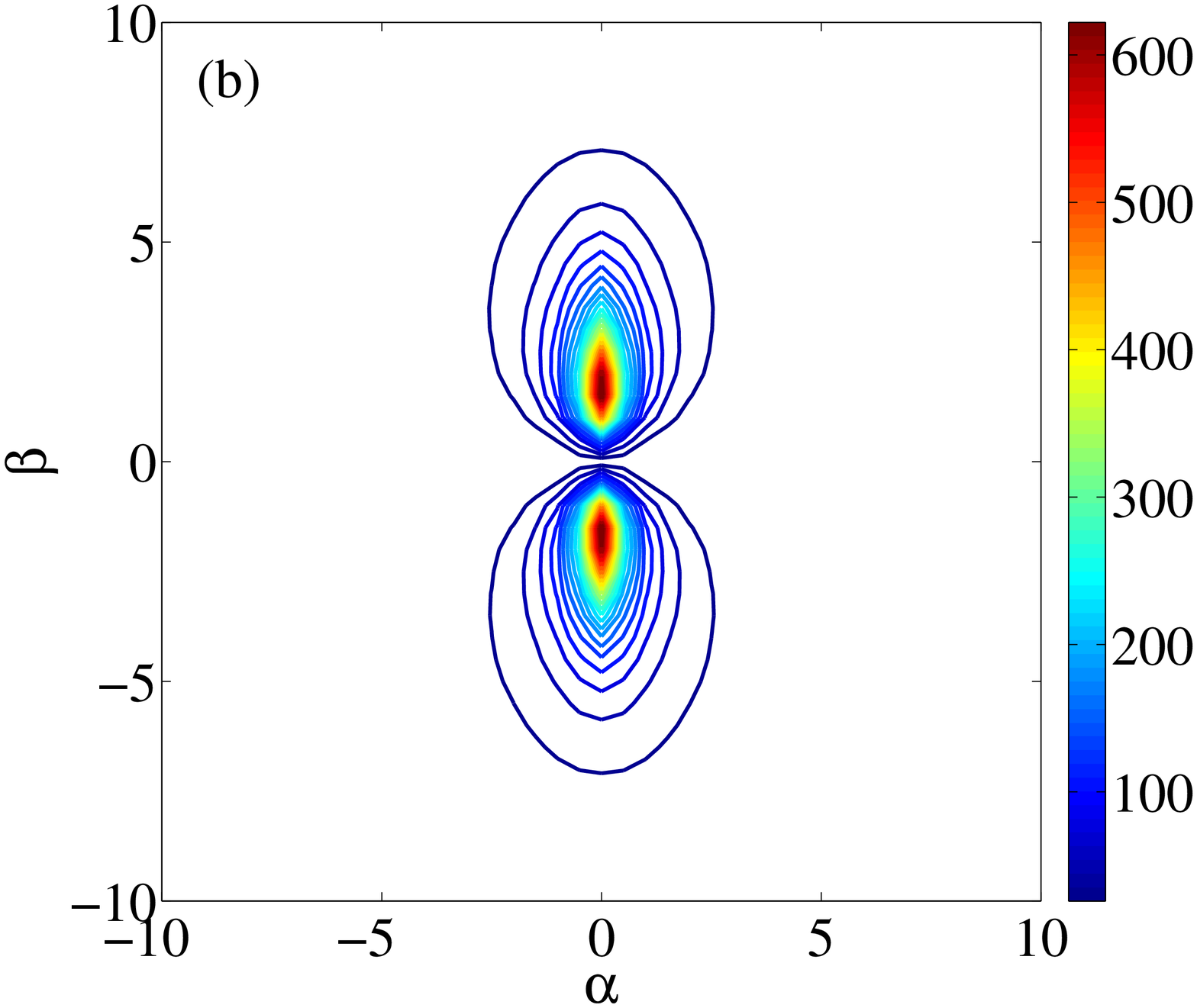}\\
  \includegraphics[width=0.4\textwidth]{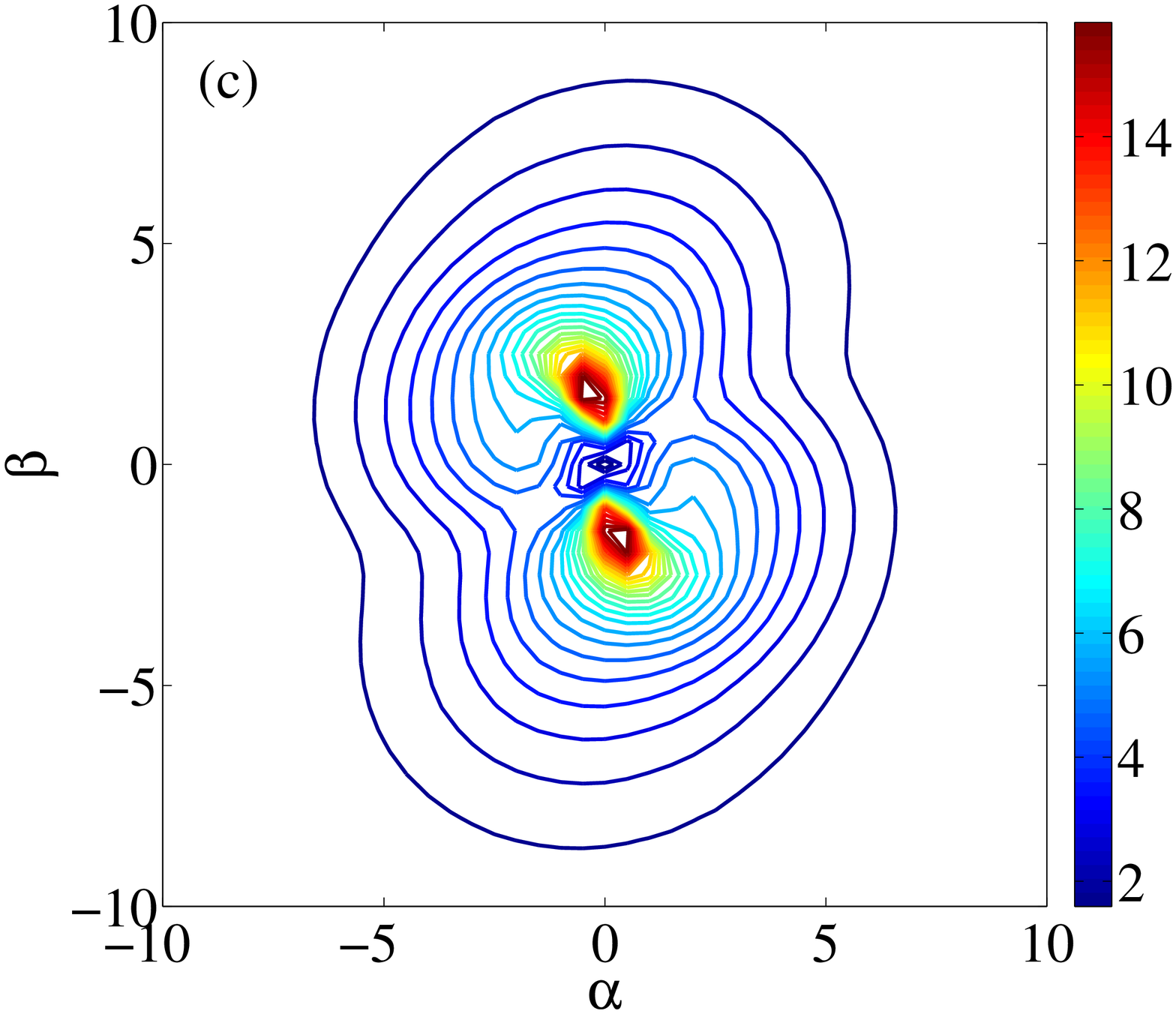}
  \includegraphics[width=0.4\textwidth]{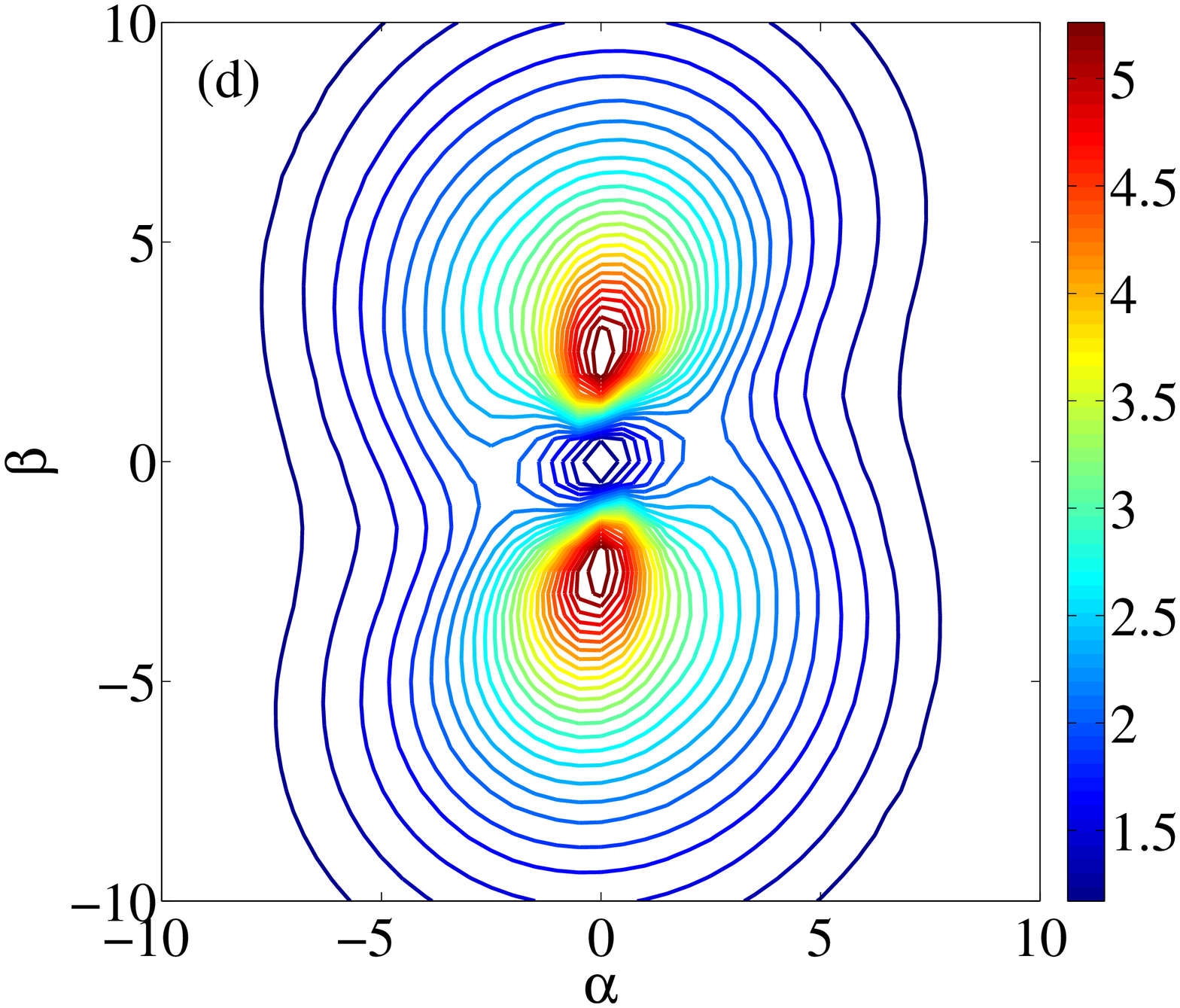}\\
  \caption{(Color online) Contour plot in $(\alpha,\beta)$ space of the global  maximum growth rate 
    $G^{m}$: (a) $\text{Re}=500,\Omega=0.05$; (b) $\text{Re}=1500,\Omega=0.050$; 
    (c) $\text{Re}=500,\Omega=20$; (d) $\text{Re}=500,\Omega=50$. 
    The size of $G^{m}$ is indicated by the color value.}
  \label{fig:alphabetaContour}
\end{figure*}

Below the neutral stability curve $\text{Re}^c(\Omega)$, the flow is linearly stable and the 
transient growth of initial perturbations may play an important role in 
the nonlinear transition to turbulence. Due to the non-normality of the governing linear operator $L$, 
pCf undergoes substantial transient growth before nonlinear interaction sets in, 
~\cite{ButlerFarrel_pra1992,ReddyHenningson_jfm1993}. 
However, the influence of the external system rotation on the transient behavior is still unknown. 
Here we employ the method presented in~\cite{ReddyHenningson_jfm1993} to compute the optimal 
transient growth and the optimal perturbations. Let us first define the physical quantities of interest, 
the spectral energy
\begin{equation}
  \hat{E}(\alpha,\beta,\text{Re},\Omega;t)=\|\hat{\textbf{v}}\|^2 = \int_{-1}^1(|\partial_y\hat{v}|^2 
  + k^2|\hat{v}|^2 + |\hat{\eta}|^2)dy
\end{equation}
and the optimal growth function
\begin{equation}
  G(\alpha,\beta,\text{Re},\Omega;t)=\sup_{\hat{\textbf{v}}(\cdot;0)\neq 0}\frac{\hat{E}(\cdot;t)}{\hat{E}(\cdot;0)} 
  = \sup_{\hat{\textbf{v}}(\cdot;0)\neq 0}
  \frac{\|\hat{\textbf{v}}(\cdot;t)\|}{\|\hat{\textbf{v}}(\cdot;0)\|}.
\end{equation}
The spectral energy measures the kinetic energy contained in the mode $(\alpha,\beta)$, while 
the optimal growth function is the maximal energy growth achievable among all possible 
initial perturbations within time $t$. As presented in~\cite{ReddyHenningson_jfm1993}, the growth function
$G(\cdot;t)$ can be directly computed by the eigenvalues and eigenvectors of the linear operator $L$.
 For simplicity of notation, we here introduce two additional functions: the maximal growth in time as 
$G^{m}(\alpha,\beta, \text{Re},\Omega)=\max_{t}G(\alpha,\beta,\text{Re},\Omega;t)$ and the global optimal growth in the 
$(\alpha,\beta)$ plane as $G^{opt}(\text{Re},\Omega)=\max_{\alpha,\beta}G^m(\alpha,\beta,\text{Re},\Omega)$. 
One property of $G^m$ is the symmetry under the transformation 
$(\alpha\to -\alpha, \beta \to -\beta)$. The maximal growth $G^m$ in 
the $\alpha$-$\beta$ plane at various \text{Re} and $\Omega$ is shown 
in Fig.~\ref{fig:alphabetaContour}, evidencing the 
symmetry with respect to the point $(0,0)$. The \text{Re}ynolds number and the rotation number are 
fixed in each case. The range in the $\alpha$-$\beta$ plane is $[-10,10]\times [-10,10]$. 
At small $\Omega$ (Fig.~\ref{fig:alphabetaContour}a), the contour plot of $G^m$ is similar 
to that in plane Couette flow~\cite{ReddyHenningson_jfm1993}, where the maximum is located very 
close to the $\beta$ axis. As $\Omega$ increases, the effect of the external rotation 
becomes non negligible and the maximum moves away from the $\beta$ axis. 
Moreover, increasing \text{Re} from 500 (Fig.~\ref{fig:alphabetaContour}a) to 1500 (Fig.~\ref{fig:alphabetaContour}b) 
results in a substantial increase of $G^m$ while 
increasing $\Omega$ from 0.05 (Fig.~\ref{fig:alphabetaContour}a) to 50 (Fig.~\ref{fig:alphabetaContour}d) 
leads to a sharp decrease in $G^m$. 
It is worthwhile to note that the modes that achieve the maximal transient growth are 
not the least unstable modes computed from the linear stability analysis.

\begin{figure}[h!]
  \centering
  \includegraphics[width=0.45\textwidth]{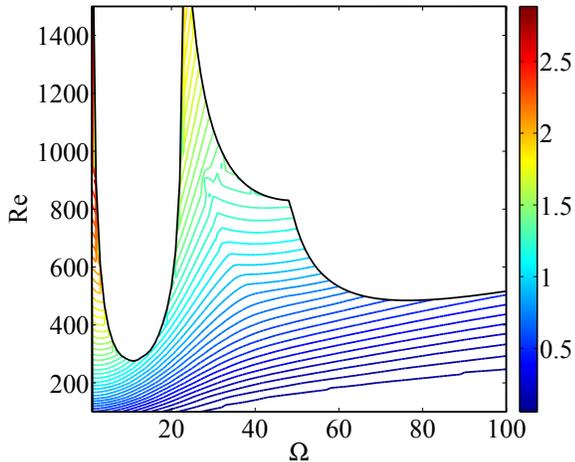}\\
  \caption{(Color online) Contour plot of the global optimal growth $G^{opt}$ in the $\Omega$-$\text{Re}$ plane. 
    The boundary (the black line) is the neutral curve from the linear stability analysis.
    The color value is on a logarithmic scale, \textit{e. g.}, 
    the value ``2'' denotes $G^{opt}=10^2$.}
  \label{fig:OmegaRe100}
\end{figure}

\begin{figure}[h!]
  \centering
  \includegraphics[width=0.45\textwidth]{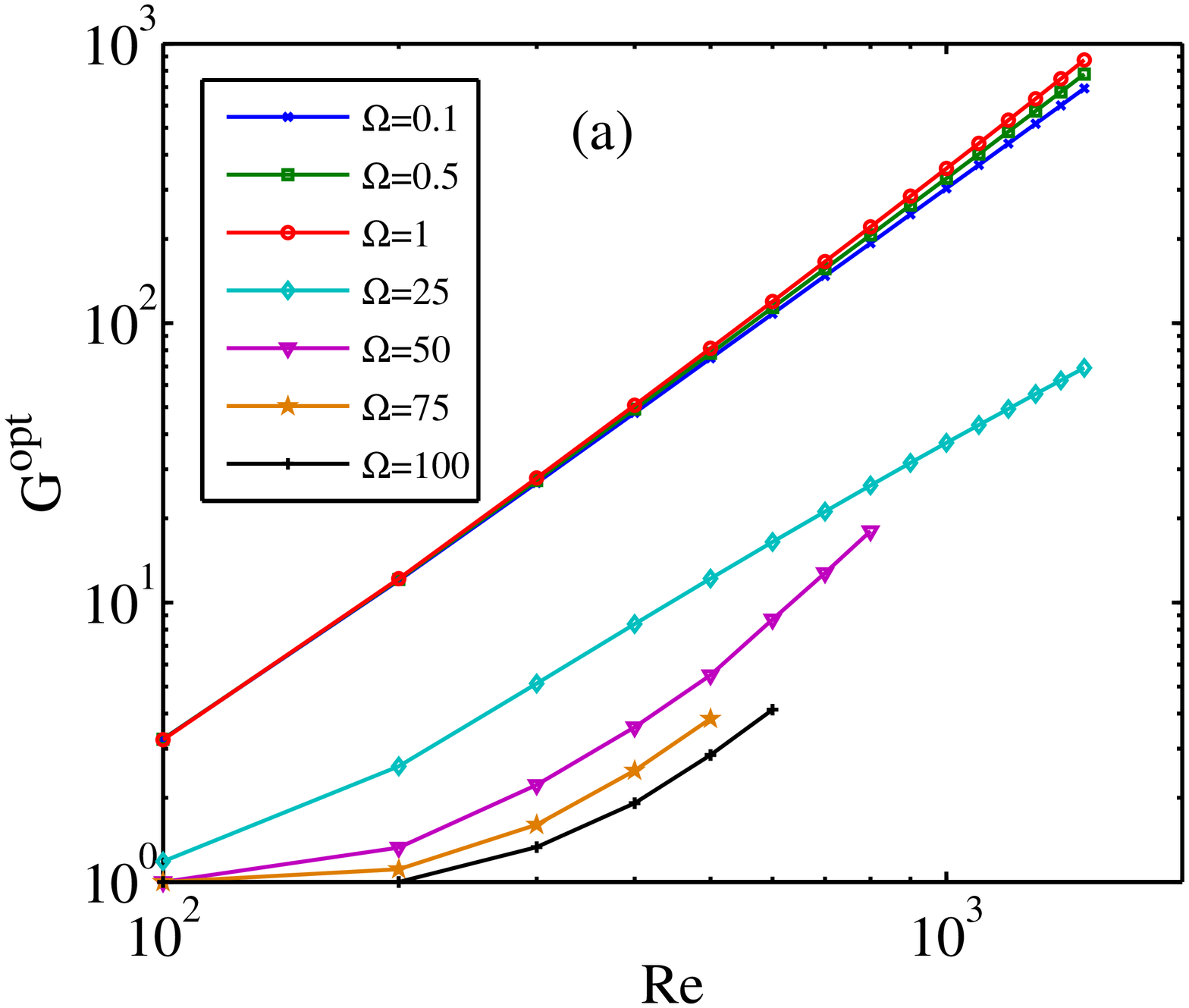}\\
  \includegraphics[width=0.45\textwidth]{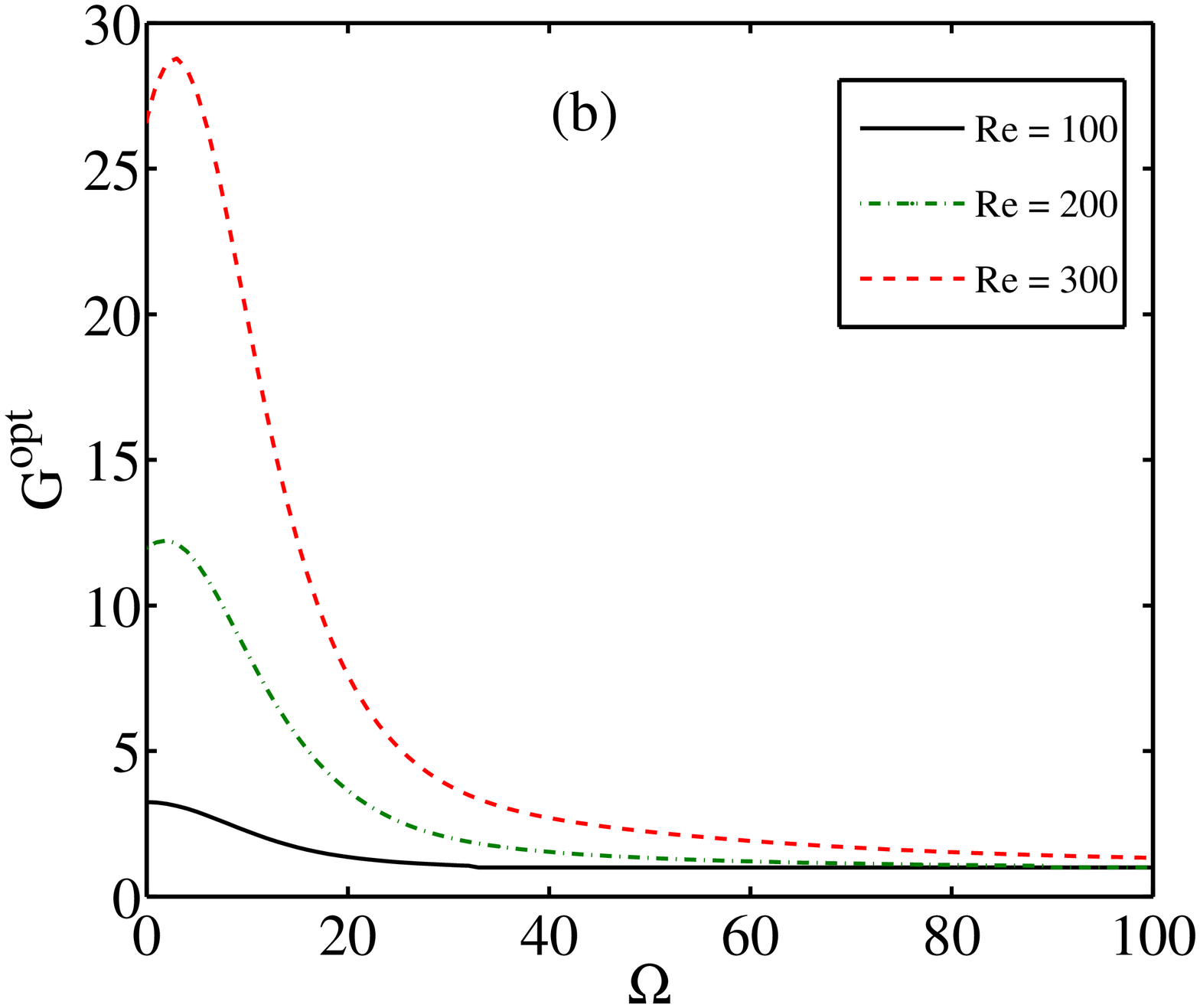}
  \caption{(Color online) Scaling of $G^{opt}$ (a) with $\text{Re}$ at different $\Omega$ and (b) with $\Omega$ at different $\text{Re}$.}
  \label{fig:gOptRe_v2}
\end{figure}

We further compute the global optimal growth function $G^{opt}(\text{Re},\Omega)$ 
in the linearly stable region in $\text{Re}\in [0,100]$ and $\Omega\in[0,1500]$. 
The search of the global maximum in the $\alpha$-$\beta$ plane is done by
the downhill simplex method~\cite{PressFlannery_2007}. Fig.~\ref{fig:OmegaRe100} 
shows the contour plot of $G^{opt}(\text{Re},\Omega)$. The highest growth 
is located in the left top region with low $\Omega$ and high \text{Re}, while the 
lowest growth is in the right bottom region with high $\Omega$ and low \text{Re}. 
Nevertheless, the middle bumps in the contour plot shows that the 
growth variation is not monotonic, whereas the non-smooth, irregular patches 
are due to the lack of sufficiently high resolution.
Quantitatively, the scaling of $G^{opt}$ with \text{Re} and $\Omega$ is shown in 
Fig.~\ref{fig:gOptRe_v2}. 
Fig.~\ref{fig:gOptRe_v2}a displays the variation of $G^{opt}(\text{Re},\Omega)$ 
with \text{Re} at fixed $\Omega$. The optimal growth scales at small $\Omega$ slightly faster 
than a power law with \text{Re}, $G^{opt}\sim \text{Re}^2$, whereas the power-law scaling disappears 
at large $\Omega$ and the transient growth becomes much smaller than the one at small $\Omega$. 
Fig.~\ref{fig:gOptRe_v2}b plots the growth $G^{opt}$ as a function of $\Omega$ when \text{Re} is fixed. 
It can be seen that the transient growth is enhanced with weak external rotation while 
it is dramatically suppressed as $\Omega>25$.

\begin{figure*}[!h]
  \centering
  \includegraphics[width=0.475\textwidth]{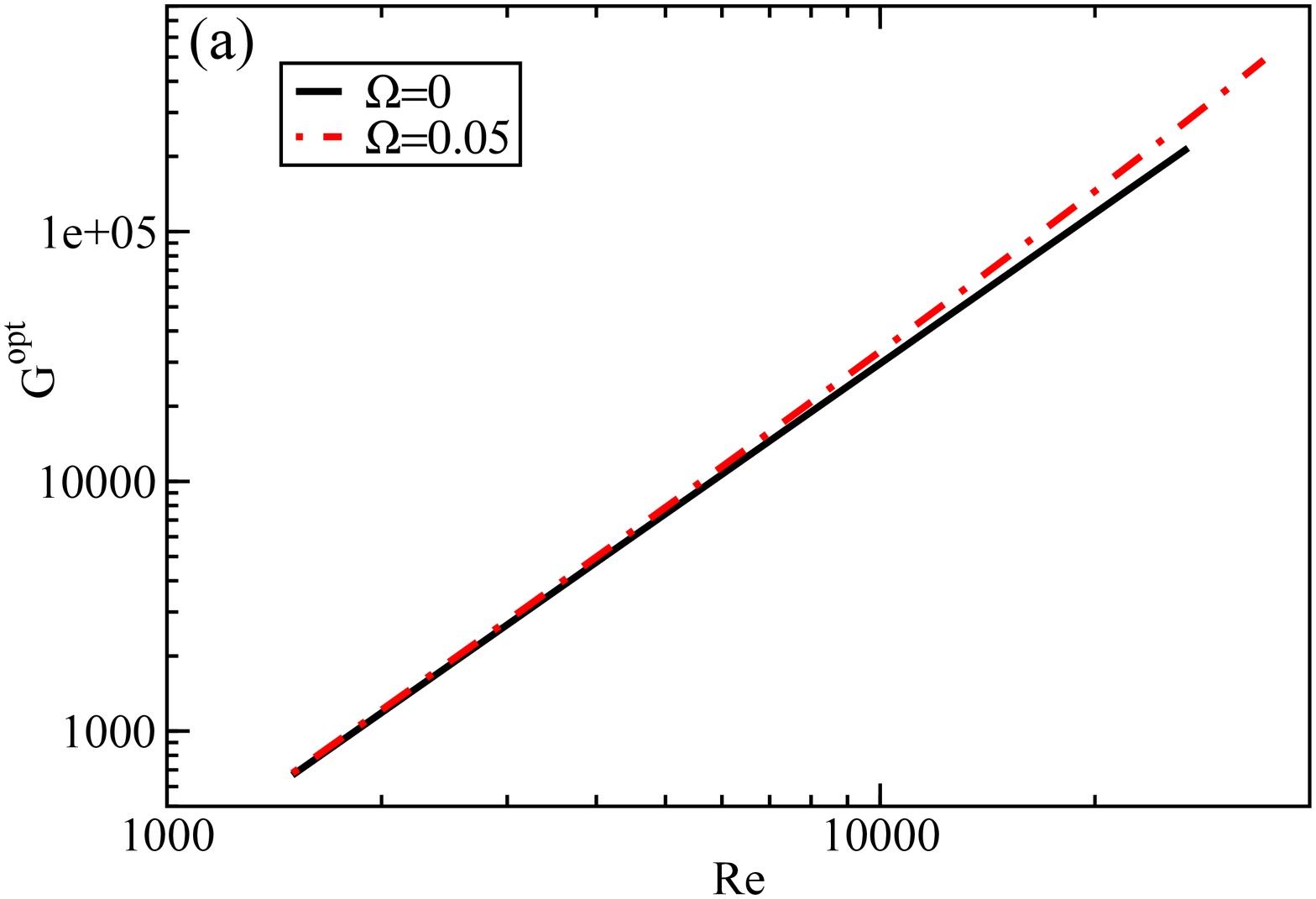}
  \includegraphics[width=0.475\textwidth]{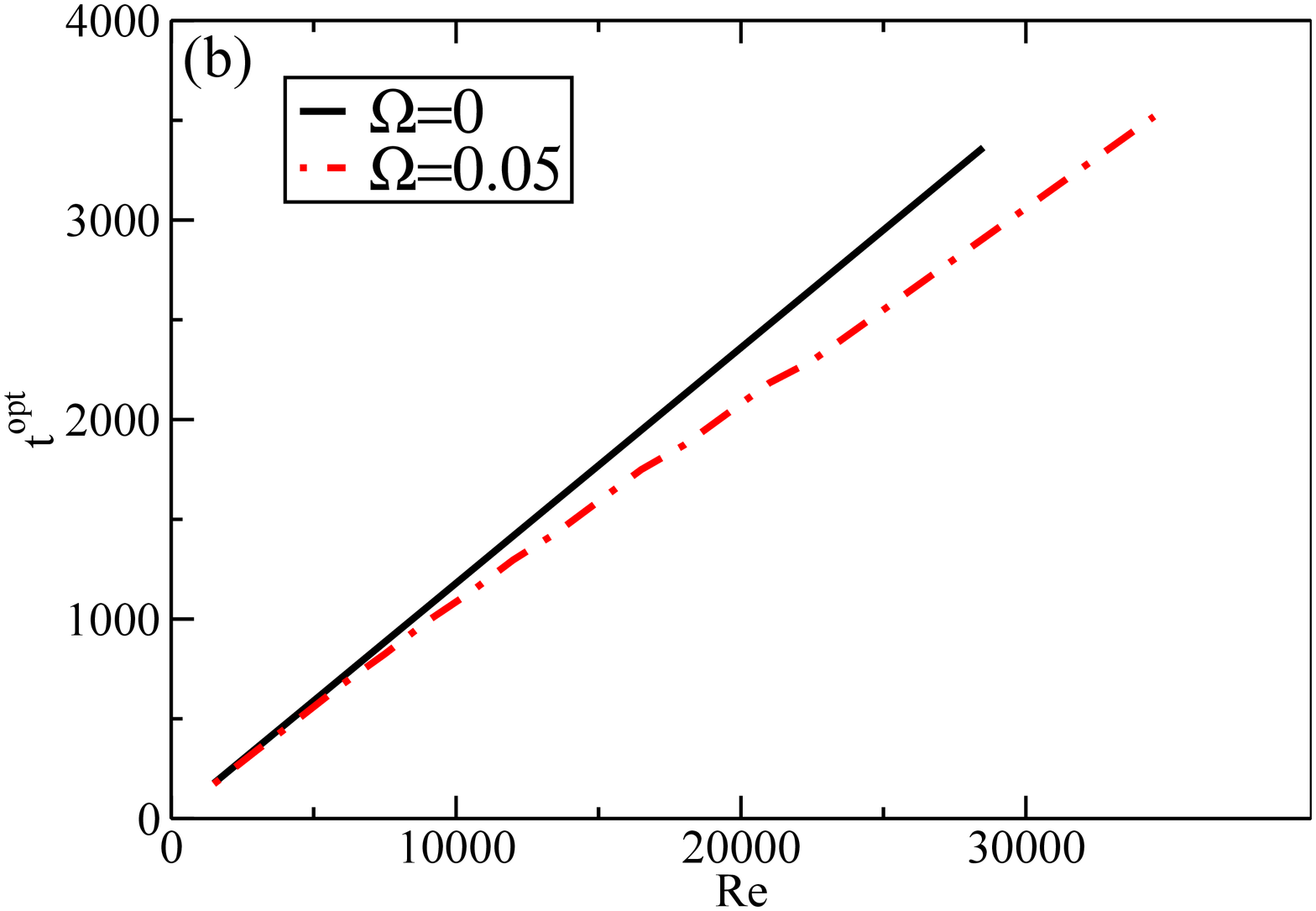}\\
  \includegraphics[width=0.475\textwidth]{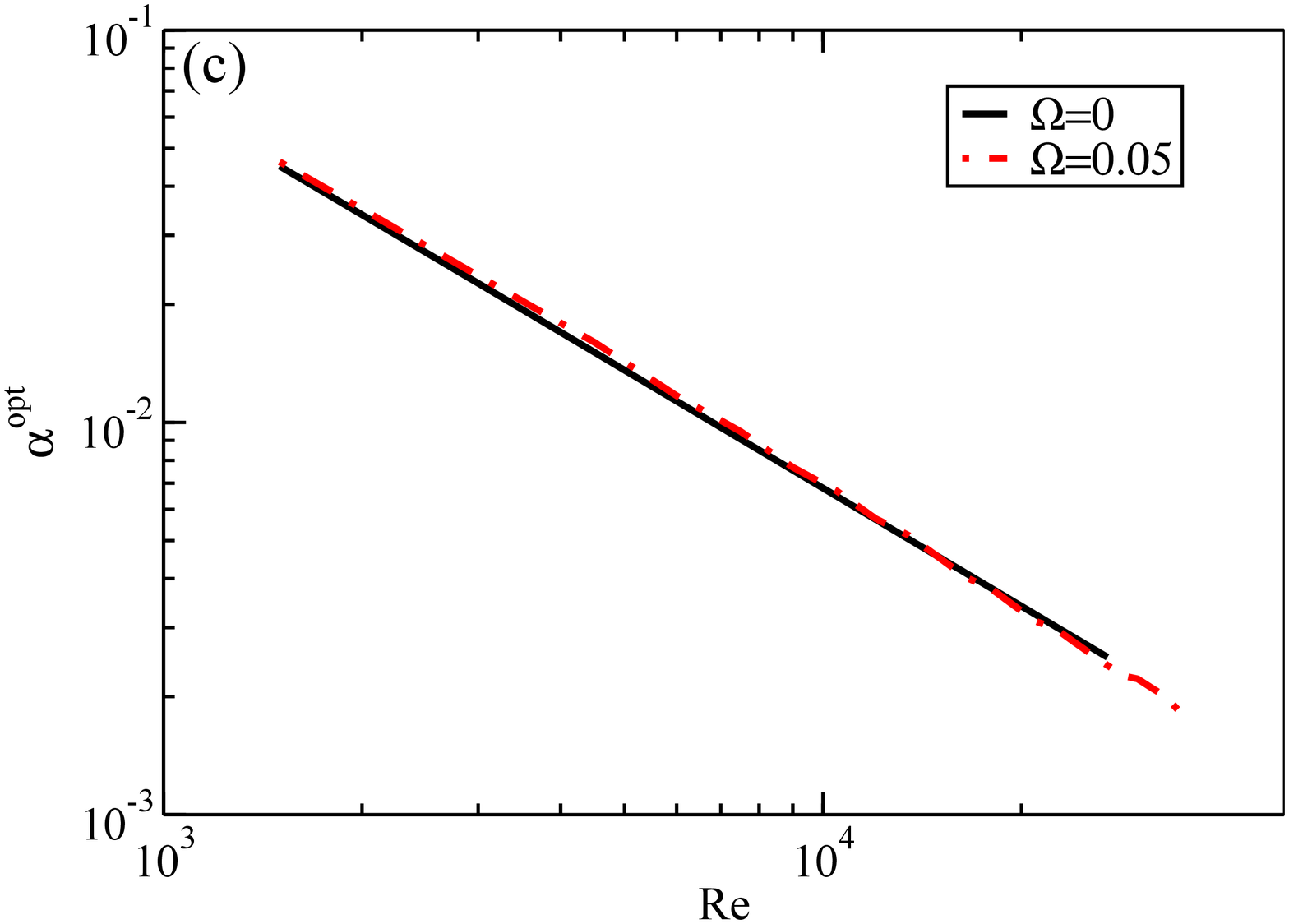}
  \includegraphics[width=0.475\textwidth]{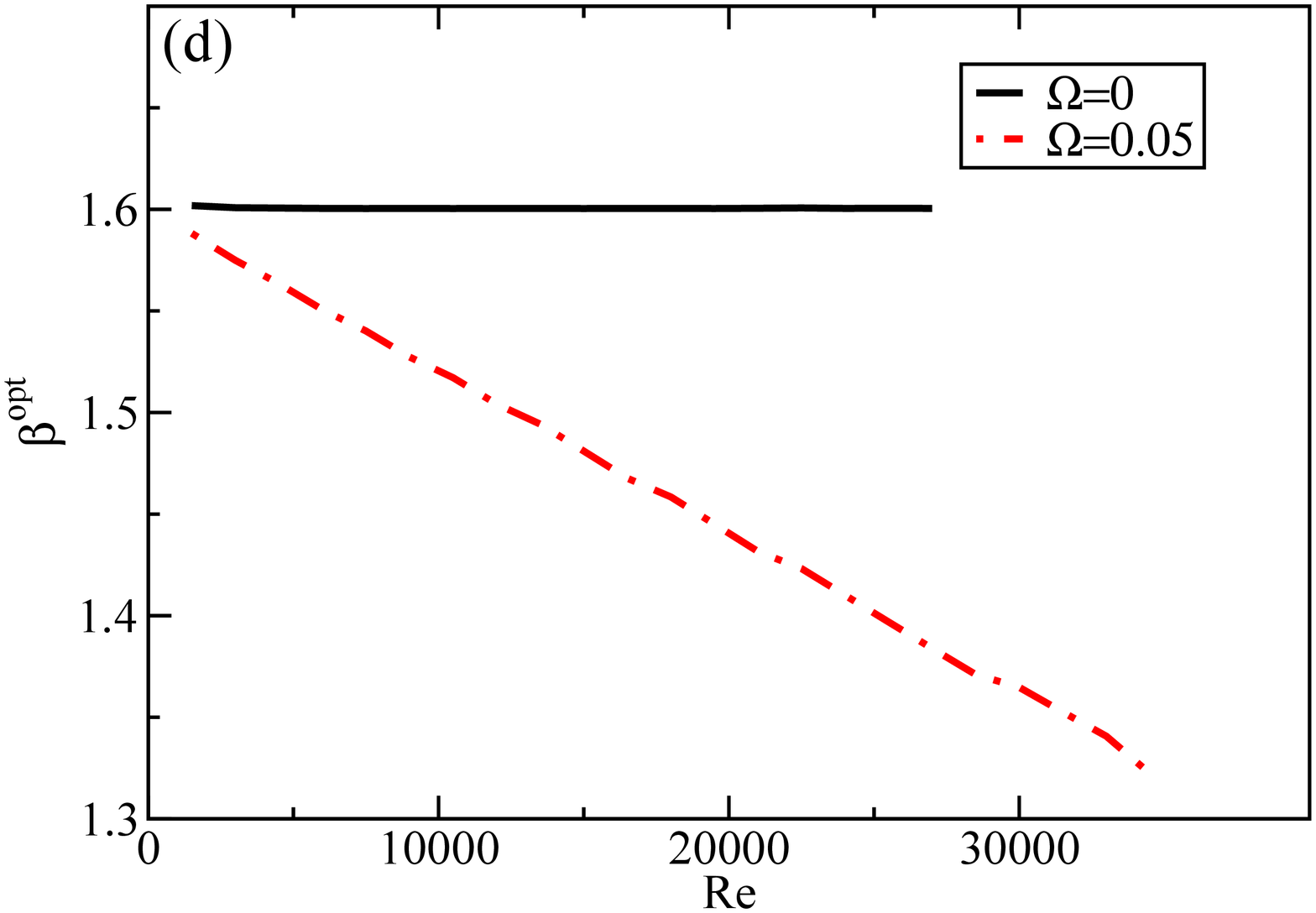}\\
  \caption{(Color online) Scaling with \text{Re}ynolds number at $\Omega =0$ and $\Omega = 0.05$. 
    (a) The global mamixum $G^{opt} \sim \text{Re}^{-2}$; 
    (b) The corresponding time where the global maximum is attained, 
    $t^{opt} = a + b\cdot \text{Re}$; (c) Wavenumber $\alpha^{opt}(\text{Re})\sim \text{Re}^{-1}$, 
    which is almost the same at different $\Omega$; (d) Wavenumber 
    $\beta^{opt}(\text{Re}) = c + d\cdot \text{Re}$, with different slopes at 
    different $\Omega$.}
  \label{fig:compare}
\end{figure*}

\begin{figure*}[h!]
  \centering
  \includegraphics[width=0.48\textwidth]{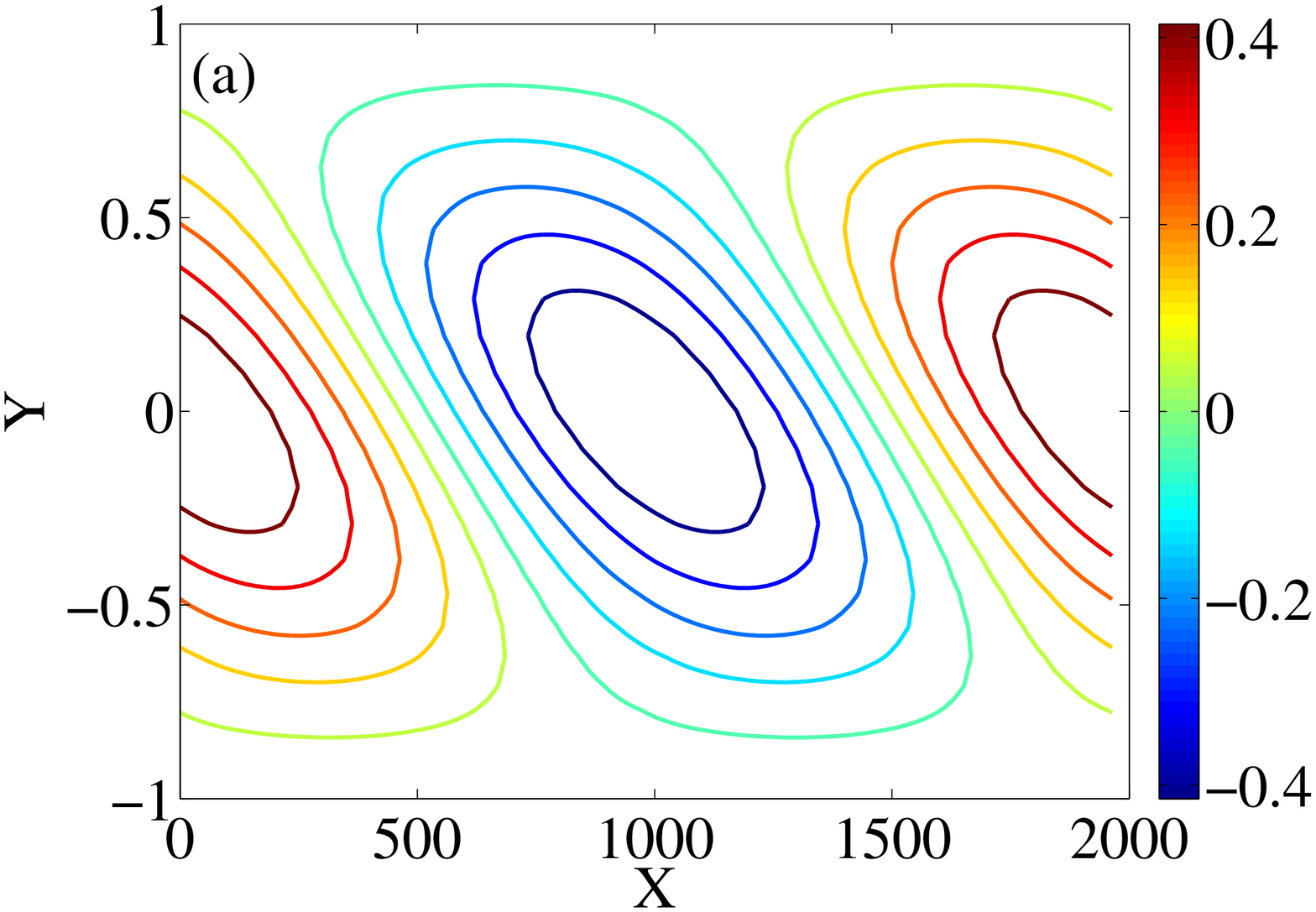}
  \includegraphics[width=0.48\textwidth]{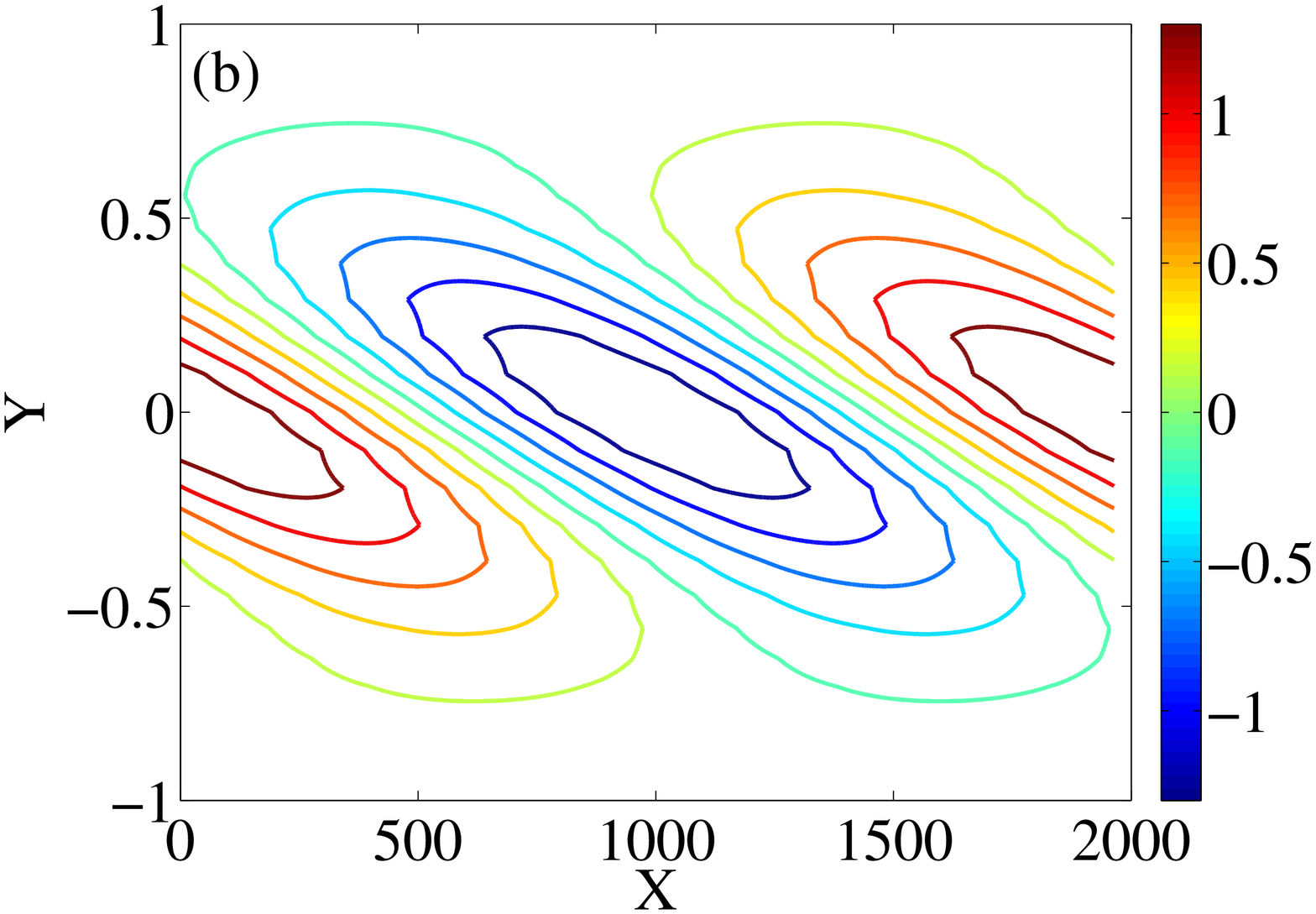}
  \caption{(Color online) Contour plot of the optimal perturbation $v^{opt}(x,y,z=0)$ 
    for (a) $\Omega = 0$ and (b) $\Omega=0.05$. The \text{Re}ynolds number 
    is $\text{Re}=20000$. The wavenumbers are the ones giving the optimal transient 
    growth.}
  \label{fig:vOpt}
\end{figure*}

The transient growth in rotating Couette flow is finally compared to 
the case without external rotation. Here in the rotating 
case we choose $\Omega=0.05$. For a plane Couette setup in G\"ottingen (Germany) 
with a gap distance $D\simeq 0.03 m$, the rotation number induced by the 
Earth's rotation $\Omega_0 \simeq 7.3\cdot 10^{-5} \cdot \sin(51.32\cdot \pi/180) 
\simeq 5.7 \cdot 10^{-5}$ is $\Omega = \frac{\Omega_0 D^2}{\nu_{H_2O}} \simeq 0.0513$, 
where the water viscosity is $\nu_{H_2O}\simeq 10^{-6} m^2/s$ at $T = 20^o$.
The \text{Re}ynolds number under investigation is in the range $\text{Re}\in[1500, 35000]$.
The results are ploted in Fig.~\ref{fig:compare}. In pCf, we have 
the optimal transient growth $G^{opt}\simeq 1.18\times 10^{-3}\text{Re}^2$ (Fig.~\ref{fig:compare}a) achieved at time 
$t^{opt}\simeq 0.117\text{Re}$ (Fig.~\ref{fig:compare}b), which agrees perfectly with the results in~\cite{ButlerFarrel_pra1992}. 
The corresponding wavenumber $\alpha^{opt}$, as shown in Fig.~\ref{fig:compare}c, 
scales as $\alpha^{opt}\sim \text{Re}^{-1}$ and $\beta^{opt}$ (Fig.~\ref{fig:compare}d) stays constant, $\beta^{opt}\simeq 1.60$. 
For the case with external rotation $\Omega=0.05$, the transient growth is slightly increased, with 
a power exponent little greater than $2.0$ and it is obtained at an earlier moment 
(see Fig.~\ref{fig:compare}b). The wavenumber $\alpha$ 
is basically the same as the case without rotation, while the wavenumber $\beta$ decreases linearly 
with \text{Re} and has a different slope at different $\Omega$. Furthermore, as shown in Fig.~\ref{fig:vOpt}, 
the optimal perturbations are both in the form of inclined roll structures. 
However, the elongated rolls in the case of $\Omega=0.05$ are slightly twisted. 

\section{Conclusion}

We presented in this paper a study of the linear stability and transient 
energy growth in rotating plane Couette flows, where the rotation axis is 
perpendicular to the planes. Such a rotating framework is of interest 
to geophysical and astrophysical flows. For example plane Couette and 
Taylor Couette experiments that are often used to study the stability of 
geophysical and astrophysical flows~\cite{JiGoodman_nature2006} 
are all exposed to the Earth's rotation. 
By linearizing the Navier-Stokes equations, we firstly computed the neutral 
stability curve dividing the linearly stable and unstable region in the 
Re-$\Omega$ parameter space. Three differnet type of instabilities are found: 
for $\Omega>20$, type I and type II instabilities which have been already known 
from the Ekman boundary layer flow and, for $\Omega<20$, type ``0'' instabilities. 
The results are consistent with the previous one reported 
in~\cite{HoffmannChen_jfm1998, PontySoward_jfm2003}.
Moreover, we found that the critical \text{Re} for $\Omega<5$ scales 
as a power law with $\Omega$, $\text{Re}^c(\Omega)\simeq 1800\cdot \Omega^{-1}$ , which 
agrees with the fact that the pCf ($\Omega=0$) is linearly stable for all \text{Re}.

Through computation of the eigenvalues and eigenfunctions of the governing linear operator $L$, 
we obtained the global optimal transient growth in the $\alpha$-$\beta$ plane 
amongst all possible initial perturbations. 
Our results show that the external rotation can have both enhancing and suppressing effects 
on the optimal transient growth.
For weak rotation, it increases the transient growth while strong rotation inhibits 
significantly the transient growth. At the rotation numbers relevant for geophysical 
applications, for example the atmospheric boundary layer, the transient growth is 
so small that linear stability analysis appears to be the appropiate tool to 
determine the stability limits of Ekman layers in the geophysical context.
At small rotation the optimal growth 
scales slightly faster than the power law $\Omega^2$ as is found in plane Couette flow.
Furthermore, the wavenumbers where the optimal 
transient growth is obtained is also different from the non-rotating case. 
The optimal wavenumber $\alpha$ stays the same, scaling as a power law $\alpha\sim \text{Re}^{-1}$, 
whereas the optimal wavenumber $\beta$ is shifted linearly with \text{Re}. 

\begin{table}[!h]
  \centering
  \begin{tabular}{l c c c r}
    \hline \hline
    place & gap $d$/mm & $\Omega$ & $\text{Re}^{c,\text{linear}}$ & $d^{\star}$/mm\\
    \hline  
    Toronto~\cite{Aydin_expFluids1991} & 58 & 0.169 & 10651 & 234.8\\
    Stockholm~\cite{Tillmark_jfm1992} & 10 & 0.006 & $3\times 10^5$ & 214.8\\
    Paris~\cite{BottinChate_epjb1998} & 7 & 0.003 & $6\times 10^5$ & 212.7\\
    Z\"urich~\cite{Krug_expFluids2012} & 31.2 & 0.052 & 34615 & 227.7\\
    \hline \hline
  \end{tabular}
  \caption{Existing experimental setups of plane Couette flows and their onset 
    of linear instability under Earth's rotation. The value $\text{Re}^c$ is
    computed according to $\text{Re}^c\simeq 1800\cdot \Omega^{-1}$, while $d^{\star}$ 
    corresponds to the gap distance beyond which the linear instability sets in 
    before the nonlinear transition to turbulence in pCf.}
  \label{tab:exppCf}
\end{table}

Mostly, the rotation of the Earth has been intuitively considered to be too weak 
to influence the experiments qualitatively. However, our results tell us that 
in the case of pCf 
the Earth's rotation does change radically the flow stability, from linearly 
stable to linearly unstable. This instability may be attributed to the inflection 
points in the base velocity profile introduced by the external rotation. 
Table~\ref{tab:exppCf} lists the existing experimental 
pCf setups and their approximate critical \text{Re}ynolds number for the linear 
instability under Earth's rotation. The value $d^{\star}$ indicates 
a reference gap distance where $\text{Re}^{c,\text{linear}}=\text{Re}^{c,\text{nonlinear}}$, 
\textit{i. e.}, the critical Reynolds number from the linear instability equals 
the one computed from nonlinear mechanism in pCf ($\sim 650$ based on the gap distance, 
see~\cite{BottinChate_epjb1998, Duguet_jfm2010, LiangHof_prl2013}).
Although the linear $\text{Re}^c$ are far beyond 
the onset of turbulence via nonlinear mechanism, 
the results provide important theoretical guidance for the design of future pCf setups. 
It may also be relevant to recent Taylor-Couette studies at \text{Re} of 
order $\mathcal{O}(10^6)$~\cite{JiGoodman_nature2006,SchartmanGoodman_aa2012,PaolettiLathrop_prl2011}, 
in that at large Re the additional component of rotation induced by the Earth's rotation 
may also cause inflection points in the base velocity profile.
Further studies on the underlying physical mechanisms will contribute to the 
understanding of shear flows in rotating frameworks.

\vspace{10pt}
L. S. appreciate the fruitful discussions with Prof. Marc Avila and Dr. Xing Wei. We acknowledge 
the research funding by Deutsche Forschungsgemeinschaft 
(DFG) under Grant No. SFB 963/1 (project A8) and the support from the Max Planck Society. 

\bibliography{./ref_ekman_couette.bib}

\end{document}